\documentclass[twocolumn,twoside]{revtex4}
\usepackage{graphicx}
\usepackage{amsmath}
\usepackage{bm}
\usepackage{fancyhdr}
\begin{document}

\title{$\bm{B\to D^\ast\ell\nu}$ at non-zero recoil}

%

\author{A. Vaquero}
\affiliation{Department of Physics and Astronomy, University of Utah, Salt Lake City, UT 84112-0830, USA}

\author{C. DeTar}
\affiliation{Department of Physics and Astronomy, University of Utah, Salt Lake City, UT 84112-0830, USA}

\author{A. X. El-Khadra}
\affiliation{Department of Physics, University of Illinois, Urbana, IL 61801-3080, USA and \\ Theoretical Physics Department, Fermi National Accelerator Laboratory, Batavia, IL 60510-5011, USA}

\author{A. S. Kronfeld}
\affiliation{Theoretical Physics Department, Fermi National Accelerator Laboratory, Batavia, IL 60510-5011, USA}

\author{J. Laiho}
\affiliation{Department of Physics, Syracuse University, Syracuse, NY 13244-1130, USA}

\author{R. S. Van de Water}
\affiliation{Theoretical Physics Department, Fermi National Accelerator Laboratory, Batavia, IL 60510-5011, USA}

\author{(Fermilab Lattice and MILC Collaborations)}
\noaffiliation

\begin{abstract}
The current status of the lattice-QCD calculations of the form factors of the $B\to D^\ast\ell\nu$ semileptonic decay is reviewed. Particular emphasis is given to the most mature calculation at
non-zero recoil coming from the Fermilab Lattice and MILC collaborations. Blinded, preliminary results for the form factors are shown, including a preliminary, but detailed error budget. The lattice results
seem to favor a large slope at small recoil, in contrast to the latest untagged results coming from the Belle collaboration. A comprehensive comparison between the latest BGL $z$~expansions of Belle, Babar, the
lattice and a joint BGL fit including lattice and Belle data is presented, and a roadmap to improve the current calculation is discussed. The current implications for $V_{cb}$ and $R(D^\ast)$ are discussed.
\end{abstract}

\maketitle

\thispagestyle{fancy}


\section{Introduction}
During the last few years the CKM~\cite{Cabibbo:1963yz, Kobayashi:1973fv} matrix element $V_{cb}$ has been at the center of a discussion regarding the unitarity triangle and the search for new physics.
According to the latest HFLAV report~\cite{Amhis:2016xyh} and subsequent updates, there is a $\approx 2\sigma$ tension between the inclusive and the exclusive determinations. In addition there is a combined
$\approx 3\sigma$ tension between the Standard Model (SM) predictions and experimental measurements in the $R(D)$-$R(D^\ast)$ plane. Recently a promising proposal found a way to reconcile the long-standing
inclusive vs exclusive discrepancy~\cite{BIGI2017441, GRINSTEIN2017359}: the CLN parametrization~\cite{Caprini:1997mu} is becoming too restrictive, given the current precision of experiments, and some of its
assumptions should be revisited. In contrast, the BGL parametrization~\cite{Boyd:1997kz} offers a model-independent alternative, free from precision errors. Once BGL is used to extrapolate the experimental data
to zero recoil, the $V_{cb}$ discrepancy disappears. The fact that until very recently most experimental results were only published in terms of the CLN parameters fueled the controversy.

Belle was the first collaboration to publish unfolded data with their tagged analysis~\cite{BelleTagged}. More recently, a second dataset coming from an untagged analysis was published~\cite{BelleUntagged},
along with their results for a BGL fit that disagrees with the inclusive result. In light of these recent developments, the Babar collaboration also decided to rerun their analysis using the BGL
parametrization, finding that the disagreement between inclusive and exclusive still stands. Thus, even though there is some evidence that the CLN parametrization~\cite{Caprini:1997mu} is not optimal, our
woes and worries about $V_{cb}$ are far from being solved.

In this context, a lattice-QCD calculation of the form factors involved in the decay at non-zero recoil is urgent and necessary. Lattice-QCD data can help settle the parametrization issue, as well as provide
reliable results in the low recoil region, which is most prone to experimental errors. We expect the current value of $V_{cb}$ to change as a result of lattice efforts. On top of that, a lattice $R(D^\ast)$
calculation would be very welcome. Although a variety of other calculations exist~\cite{Fajfer:2012vx,Bernlochner:2017jka,Bigi:2017jbd,Jaiswal:2017rve}, none of them comes from lattice gauge theory, the
only first-principles, non-perturbative tool available to tackle QCD.

In this work we present the current status of the ongoing lattice-QCD calculations of the $B\to D^\ast\ell\nu$ semileptonmic decay form factors. We emphasize the preliminary results coming from the most mature
calculation, and we compare the current available experimental results with the preliminary predictions coming from the lattice.

\section{Notation and definitions}
The Standard Model prediction for the differential rate for exclusive $B \to D^* \ell \nu$ decay can be written in terms of the recoil parameter $w = v_{D^\ast}\cdot v_B$,
\begin{widetext}
\begin{equation}
\frac{d\Gamma}{dw}\left(B\to D^\ast\ell\nu\right) = \frac{G_F m_B^5}{48\pi^2}\left(1-r^2\right) \sqrt{w^2-1}\,\chi(w)\left|\eta_{\rm{EW}}\right|^2\left|V_{cb}\right|^2\left|\mathcal{F}(w)\right|^2,
\label{dRate}
\end{equation}
\end{widetext}
where $v_X = p_X/m_X$ are the four velocities of the $B$ and $D^\ast$ mesons, $\eta_{\rm{EW}}$ is a correction factor that accounts for electroweak effects, $r = m_{D^\ast}/m_B$, $\mathcal{F}(w)$ is a
function that represents the probability amplitude, to be calculated in theory, and $\chi(w)$ gathers all the remaining kinematic factors. The function $\mathcal{F}$ can be expressed in terms of the helicity
amplitudes $H_{\pm,0}$ as,
\begin{widetext}
\begin{equation}
\chi(w)\left|\mathcal{F}(w)\right|^2 = \frac{1 - 2wr + r^2}{12m_Bm_{D^\ast}(1-r)^2} \left(H_0^2(w) + H_+^2(w) + H_-^2(w)\right).
\end{equation}
The helicity amplitudes, in turn, depend on the $h_X(w)$ form factors, motivated by heavy quark effective theory (HQET),
\begin{align}
H_0  (w) =& \frac{\sqrt{m_Bm_{D^\ast}}}{1 - 2wr + r^2}(w+1)\left[(w-r)h_{A_1}(w) - (w-1)(rh_{A_2}(w) + h_{A_3}(2))\right], \\
H_\pm(w) =& \sqrt{m_Bm_{D^\ast}}(w+1)\left(h_{A_1}(w) \pm \sqrt{\frac{w-1}{w+1}} h_V(w)\right).
\end{align}

The form factors are defined following the standard decomposition of the matrix elements of the $V - A$ weak current that mediates the transition,
\begin{align}
\frac{\left\langle D^\ast(p_{D^\ast}, \epsilon^\nu)\right|\mathcal{V}^\mu\left|B(0)\right\rangle}{2\sqrt{m_Bm_{D^\ast}}} =& \frac{1}{2}\epsilon^\ast_\nu\varepsilon^{\mu\nu}_{\sigma\rho}
v_{D^\ast}^\sigma v_B^\rho\,h_V(w),\label{vecCor} \\
\frac{\left\langle D^\ast(p_{D^\ast}, \epsilon^\nu)\right|\mathcal{A}^\mu\left|B(0)\right\rangle}{2\sqrt{m_Bm_{D^\ast}}} =& \frac{i}{2}\epsilon^\ast_\nu\left[g^{\mu\nu} (1+w)h_{A_1}(w) -
v_B^\nu\left(v_B^\mu h_{A_2}(w) + v_{D^\ast}^\mu h_{A_3}(w)\right)\right].\label{axCor}
\end{align}
\end{widetext}

Up to very recently, only zero-recoil ($w=1$) results were available for the decay amplitude $\mathcal{F}(w)$ in (\ref{dRate}). HQET calculations can access only the zero recoil region because, as soon as $w$
deviates from one, non-perturbative effects become important, and we don't know how to calculate them without resorting to lattice-QCD. On the other hand, lattice-QCD calculations at non-zero recoil are
extremely complicated, and had not been tackled until the last years.

Experiments measure the differential decay rate~(\ref{dRate}), but since the kinematic factor is proportional to $\sqrt{w^2 -1}$, it is not feasible to measure it directly at zero recoil. Hence, we must rely on
extrapolations in order to extract $V_{cb}$ from a joint effort coming from experiment and theory. The most popular parametrizations for this decay are BGL~\cite{Boyd:1997kz} and CLN~\cite{Caprini:1997mu}. The
parametrizations are explained in section~\ref{Parms}.

\section{Status of the lattice calculation}
\begin{figure}[h]
\centering
\includegraphics[width=80mm]{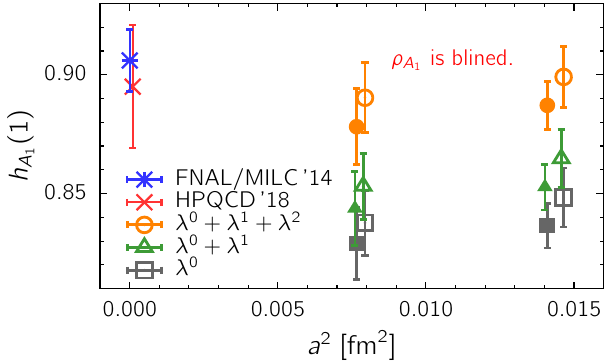}
\caption{Status of the zero recoil LANL/SWME calculation. Here $\lambda \sim \Lambda_{\rm{QCD}} / 2m_c$ and $\lambda^n$ refers to the highest HQET term included in the matching procedure. Plot taken from the
	 slides of the KEK-FF2019 meeting~\cite{KKOK}, courtesy of the authors.} \label{OKplot}
\end{figure}

Lattice-QCD allows for a direct calculation of the form factors defined in eqs.~(\ref{vecCor},\ref{axCor}) at zero and non-zero recoil. Currently only zero-recoil calculations are available, where only
$h_{A_1}$ contributes to the decay amplitude. The Fermilab Lattice and MILC collaborations~\cite{Lattice:2015rga} use the asqtad ($a-$squared, tadpole improved) action for the light sea and valence quarks, and
the Fermilab effective action for the heavy quarks. It employs 15 ensembles with lattice spacings ranging from $0.045$ fm up to $0.15$ fm, and pion masses from $500$ MeV down to $180$ MeV. Their result is
$\mathcal{F}(1)=0.906\pm0.004_{\textrm{Stat}}\pm0.012_{\textrm{Sys}}$. The HPQCD collaboration \cite{Harrison:2017fmw} has published results using 8 ensembles with the HISQ (Highly Improved Staggered Quark)
action for the sea and valence quarks, and using NRQCD for the bottom quark. Here the sea includes charm. They use lattice spacings ranging from $0.15$ fm to $0.09$ fm and pion masses from $320$ MeV up to the
physical value. Their final result is $\mathcal{F}(1)=0.895\pm0.010_{\textrm{Stat}}\pm0.024_{\textrm{Sys}}$. Even though HPQCD uses a superior light quark action and physical pion mass ensembles, they obtain
larger errors due to the mismatch between the bottom and the charm quark regularization; this mismatch introduces larger renormalization errors than in the Fermilab/MILC result. A second calculation by HPQCD
focuses on $B_s\to D_s^\ast\ell\nu$~\cite{HPQCD2}. It also uses the HISQ regularization for the bottom quark, but at an unphysical mass, requiring an extrapolation to reach the right value. Nevertheless, the
use of the same actions for the $B$ and the $D^\ast$ results in a much more accurate value than in their previous approach that used NRQCD. From their result for $\mathcal{F}_s(1)$ they extract $\mathcal{F}(1)$
by using the $\mathcal{F}(1)/\mathcal{F}_s(1)$ ratio computed in~\cite{Harrison:2017fmw}. Their final result is $\mathcal{F}(1) = 0.914\pm 0.024$, in agreement with their previous result and with the
Fermilab/MILC calculation. The fact that these independent calculations using different regularizations for the light and the heavy quarks --with totally different systematic errors-- agree, is an indication
of the reliability of the lattice calculations. There is an ongoing fourth calculation coming from the SWME and LANL collaborations~\cite{Bhattacharya:2018gan} based on the Oktay-Kronfeld action \cite{OKref}
for the bottom and charm quarks. Fig.\ref{OKplot} shows their current preliminary results, along with the previous results of HPQCD and Fermilab/MILC.

\begin{figure}[h]
\centering
\includegraphics[width=80mm]{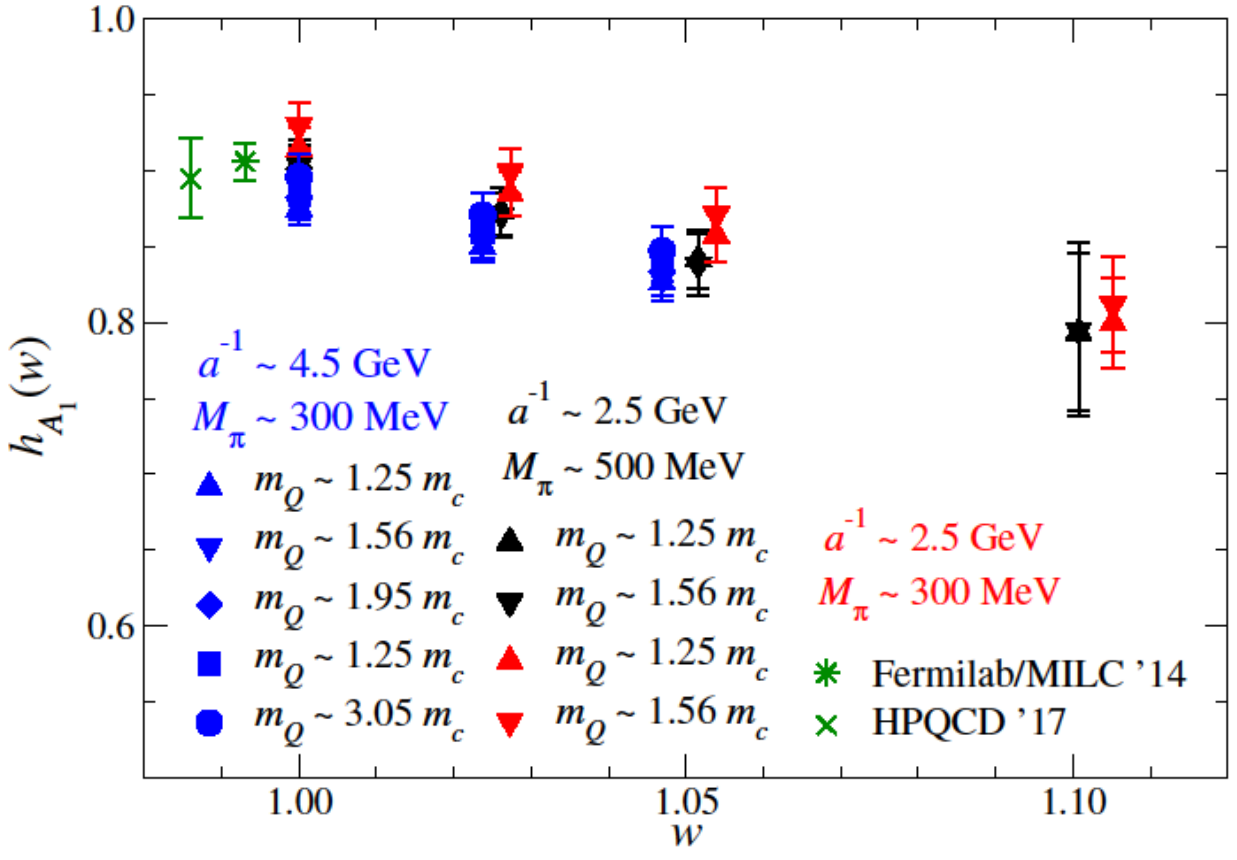}
\caption{Preliminary lattice results of the non-zero recoil JLQCD calculation for $h_{A_1}$. The collaboration is calculating the four relevant form factors. Taken from the slides of the KEK-FF2019 meeting
         \cite{KKJLQCD}, courtesy of the authors.} \label{JLQCD}
\end{figure}

\begin{figure}[h]
\centering
\includegraphics[width=80mm]{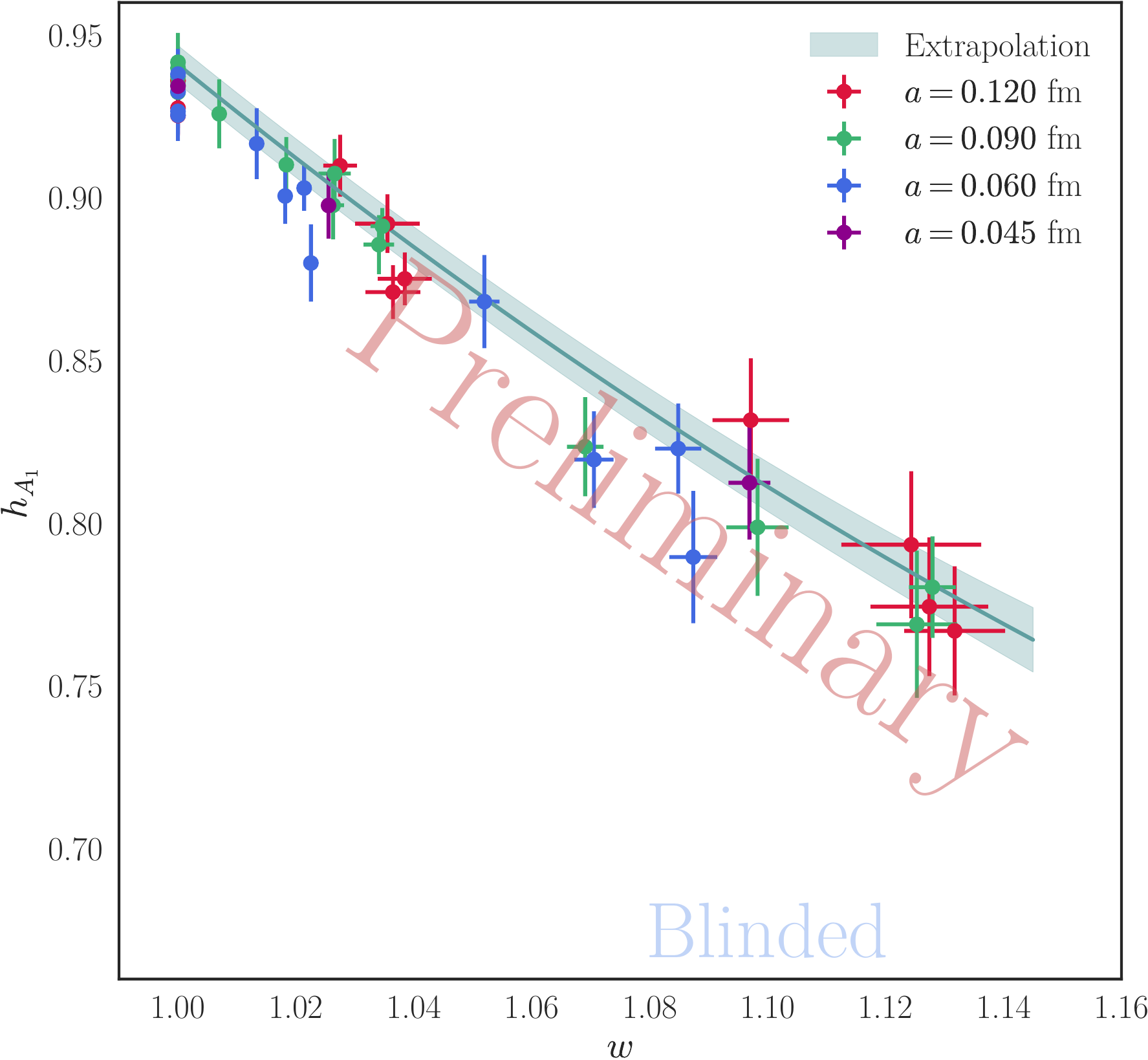}
\caption{Blinded, preliminary results of the Fermilab/MILC calculation for $h_{A_1}$. The points represent the lattice data and the band is the final result extrapolated, to the continuum.} \label{FMILC1}
\end{figure}

\begin{figure}[h]
\centering
\includegraphics[width=80mm]{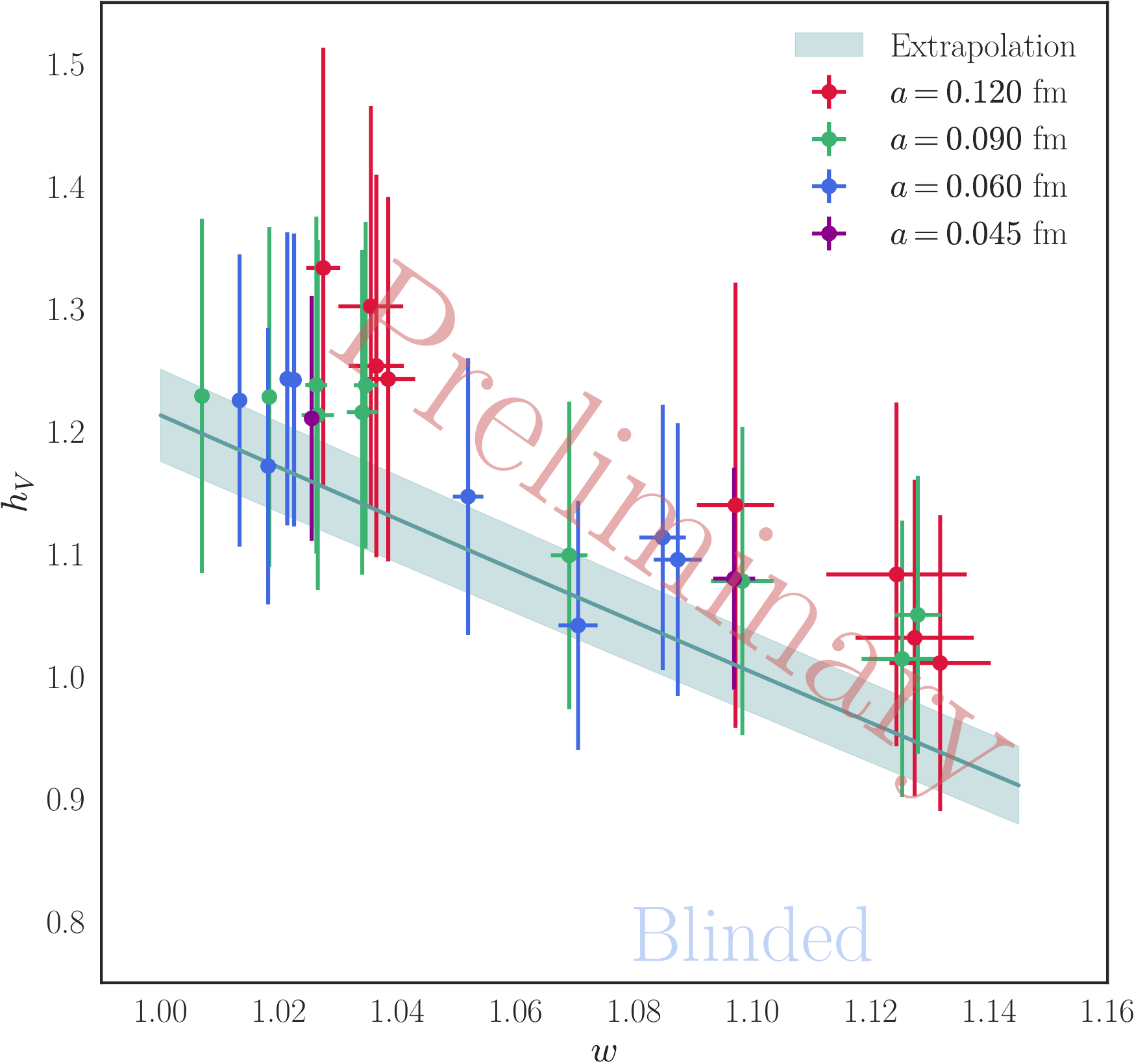}
\caption{Blinded, preliminary results of the Fermilab/MILC calculation for $h_V$. The points represent the lattice data and the band is the final result extrapolated, to the continuum.    } \label{FMILC2}
\end{figure}

\begin{figure}[h]
\centering
\includegraphics[width=80mm]{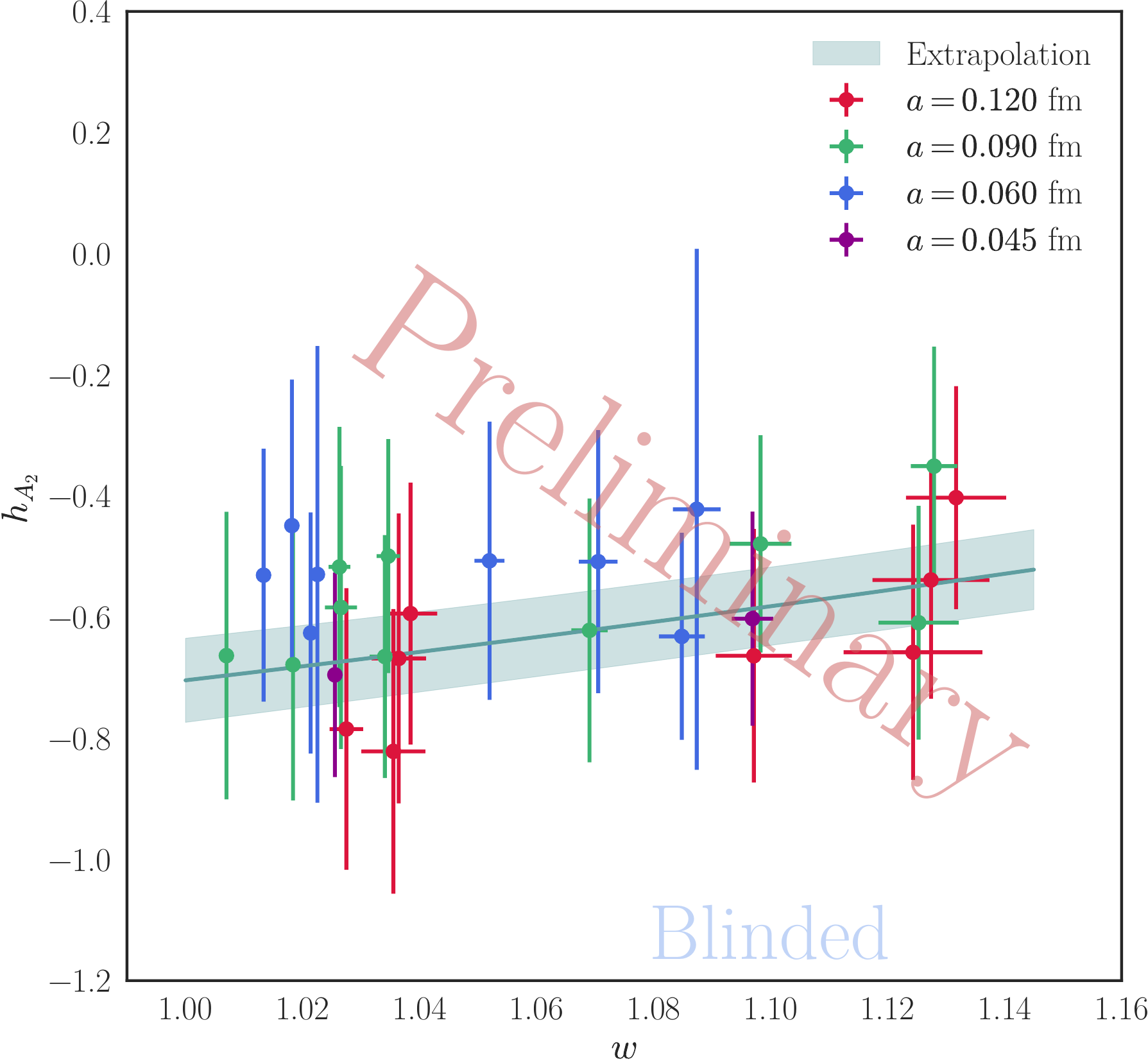}
\caption{Blinded, preliminary results of the Fermilab/MILC calculation for $h_{A_2}$. The points represent the lattice data and the band is the final result extrapolated, to the continuum.} \label{FMILC3}
\end{figure}

\begin{figure}[h]
\centering
\includegraphics[width=80mm]{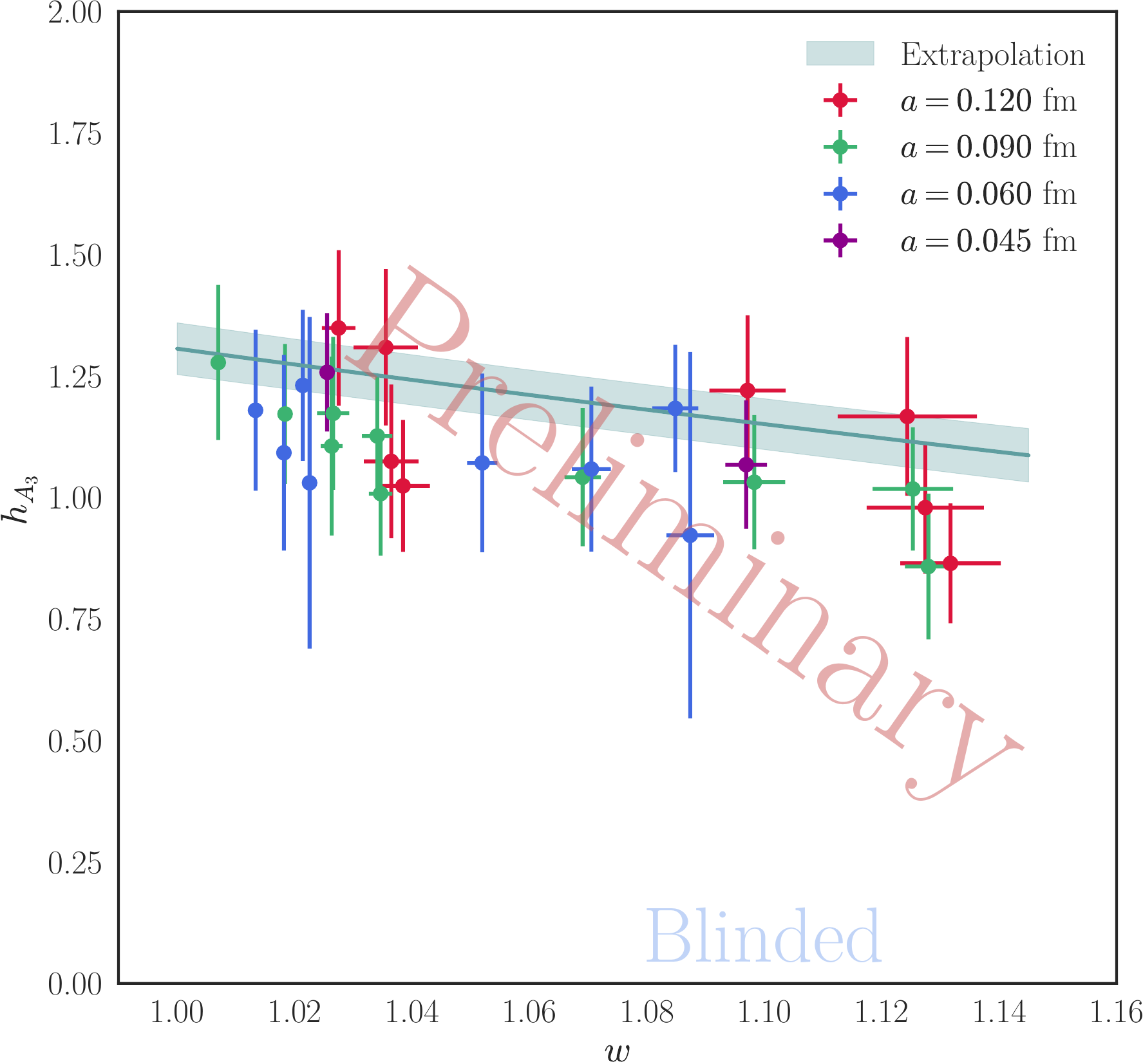}
\caption{Blinded, preliminary results of the Fermilab/MILC calculation for $h_{A_3}$. The points represent the lattice data and the band is the final result extrapolated, to the continuum.} \label{FMILC4}
\end{figure}

For non-zero recoil, the calculation becomes much more involved. As shown in Eqs.~(\ref{vecCor}) and~(\ref{axCor}), there are four different form factors to compute. One should expect that the correlations
among them are very strong, because they share underlying gauge configurations, propagators and three-point functions in their calculations. Currently there are no calculations available, but two efforts are
underway: the JLQCD collaboration is working on a calculation based on domain-wall fermions~\cite{Kaneko:2018mcr}. They include the charm quark in the sea, and the bottom quark has an unphysical mass. Then
they extrapolate to the correct bottom mass. They have presented status updates in previous conferences (see Fig.~\ref{JLQCD}). The other calculation comes from Fermilab/MILC
\cite{Aviles-Casco:2017nge,Aviles-Casco:2019vin}, and it is in its final stages of development. We use the same ensembles and regularization as in our previous zero recoil calculation, and we publish regular
updates on our analysis. Preliminary, blinded results for our form factors are shown in Figs.~\ref{FMILC1}-\ref{FMILC4}. The bands here represent the final result in the continuum, after the chiral-continuum
limit has been taken. The four form factors were fitted at the same time with $p=0.36$.

\subsection{Preliminary error budget}
Table~\ref{eBudget} gathers the preliminary error budget of the Fermilab/MILC calculation for the different form factors. The errors depend on the recoil parameter, and the table shows the values at
$w\sim 1.10$. We have a well defined roadmap to reduce these errors in three steps, and this calculation is the first of these steps.
\begin{table}
\caption{Preliminary error budget at $w\sim 1.10$.\label{eBudget}}
\begin{ruledtabular}
\begin{tabular}{l|c|c|c|c}
Source                 & $h_V\,(\%)$ & $h_{A_1}\,(\%)$ & $h_{A_2}\,(\%)$ & $h_{A_3}\,(\%)$ \\
\hline
Statistics             &     $1.1$   &     $0.4$       &      $4.9$      &      $1.9$      \\
Isospin effects        &     $0.0$   &     $0.0$       &      $0.6$      &      $0.3$      \\
$\chi$PT/cont. extrap. &     $1.9$   &     $0.7$       &      $6.3$      &      $2.9$      \\
Matching               &     $1.5$   &     $0.4$       &      $0.1$      &      $1.5$      \\
Heavy quark discr.     &     $1.4^*$ &     $0.4^*$     &      $5.8^*$    &      $1.3^*$
\end{tabular}
\footnotesize
$^*$Estimate, currently subject of intensive study.
\end{ruledtabular}
\end{table}
The main source of systematic errors currently come from the chiral-continuum extrapolation, which uses EFTs (Effective Field Theories) to take the lattice data to the continuum limit, where the lattice spacing
becomes zero, and takes the quark masses to their physical values. The use of EFTs allow us to control the systematic errors and to have a good ansatz for the extrapolation. We plan to reduce these errors 
dramatically in the second step of our roadmap, by using HISQ quarks for the light sector at smaller lattice spacings and lower pion masses, reaching physical pions in some ensembles. We plan to keep using
the Fermilab action for the charm and the bottom quarks in this first iteration.

The next important source of error comes from the discretization of the heavy quarks. The large mass of the heavy quarks makes them problematic to simulate on the lattice: since their Compton wavelength
is extremely small, they require an accordingly small lattice spacing to be simulated properly, otherwise one incurs large discretization errors. Effective heavy quark actions are used to mitigate these
effects, and in our case we employ the Fermilab action, which removes tree-level discretization effects. But still, the remaining errors constitute a large component of our current error budget. In the
third and final step of our roadmap, we plan to replace the heavy quark action with HISQ fermions as well, and simulate the bottom quark as if it were a light quark. This improvement is expected to removed most
of the discretization errors in our calculation.

The third large source of errors is the renormalization and the matching. Currents on the lattice do not have the same symmetries as currents in the continuum; therefore, they require proper renormalization.
In our calculation we compute ratios of three-point functions, where most of the renormalization factors cancel, but we must compute the remainder using perturbation theory. Currently our calculations allow
us to compute these factors at zero recoil, so we add a conservative error in order to take into account that our form factors are evaluated at non-zero recoil. This error turns out to be sizeable. Our final
iteration of the analysis with HISQ heavy quarks will use fully non-perturbative renormalization, which is expected to reduce these errors to negligible values.

The other source of systematic errors, isospin breaking, comes from the fact that in our lattice simulations we simulate with degenerate up and down quarks. This error is small compared with the other
contributions.

Once the form factors are calculated on the lattice, one can use any existing parametrization to extrapolate to the whole kinematic range with improved precision. Since the lattice is accurate only in the
small recoil region ($w\lesssim 1.1$), the parametrizations provide extremely useful information to constrain the form factors.

\section{The $\bm{z}$~expansion and $\bm{V_{cb}}$\label{Parms}}
The Boyd-Grinstein-Lebed (BGL) parametrization~\cite{Boyd:1997kz} is built upon very general arguments. It is based on the fact that one can constrain the amplitude of the production of a pair $H_b \bar{H}_c$
from a virtual $W$ boson with $q^2 > m_B^2 + m_{D^\ast}^2$, and then propagate the constraints back to the semileptonic region, where $m_\ell^2 < q^2 < m_B^2 - m_{D^\ast}^2$, by using analyticity. The process
results in bounds for the form factors, which are expressed in a general form,
\begin{equation}
F(z) = \frac{1}{B_F(z)\phi_F(z)}\sum_{j=0}^\infty a^F_j z^j,
\label{ffBLG}
\end{equation}
where $F(z)$ is a particular form factor, $B_F(z)$ is the so-called Blaschke factor, which gathers the dependence of the poles relevant to our form factor $F$, $\phi_F(z)$ is called the outer function, and must
be computed in the theory, and $z$ is a small variable coming from a comformal transformation of the recoil parameter. In the most common case, $z$ ranges from $0$ at zero recoil to its maximum value,
\begin{equation}
z = \frac{\sqrt{w+1} - \sqrt{2}}{\sqrt{w+1} + \sqrt{2}},
\label{zParm}
\end{equation}
but Babar prefers to set $z=0$ at the point $t_0=(m_B + m_{D^\ast})(\sqrt{m_B} - \sqrt{m_{D^\ast}})^2$, a particular choice that gives a symmetric range around zero.

The unitarity constraints in the BGL parametrization are applied to the coefficients $a^F_j$ of the expansion. Given a set of form factors $\mathcal{S}$ sharing parity and spin quantum numbers, the constraint
reads
\begin{equation}
\sum_{F\in\mathcal{S}}\sum_{j=0}^\infty (a^F_j)^2 \leq 1
\end{equation}
These are known as the weak unitarity constraints. One can further constrain the coefficients of the expansion by considering several decays related by crossing symmetry. The new constraints are known as the
strong unitarity constraints.

The Caprini-Lellouch-Neubert (CLN) parametrization~\cite{Caprini:1997mu} is based on the same principles as the BGL parametrization, but it builds in the strong unitarity constraints and uses some inputs based
on information in 1997 to remove parameters from the more general expansion. In the end the authors present a simplified polynomial expansion,
\begin{align}
h_{A_1}(z) =& h_{A_1}(1)\left(1-8\rho^2z + \right. \nonumber \\
            & \left.(53\rho^2-15)z^2 - (231\rho^2 - 91)z^3\right), \\
R_1(w)     =& R_1(1) - 0.12(w-1) + 0.05(w-1)^2, \\
R_2(w)     =& R_2(1) + 0.11(w-1) - 0.06(w-1)^2,
\end{align}
where $z$ is given by Eq.~(\ref{zParm}).

\section{Comparison between Lattice-QCD and Experiment}
\begin{figure}[t]
\centering
\includegraphics[width=80mm]{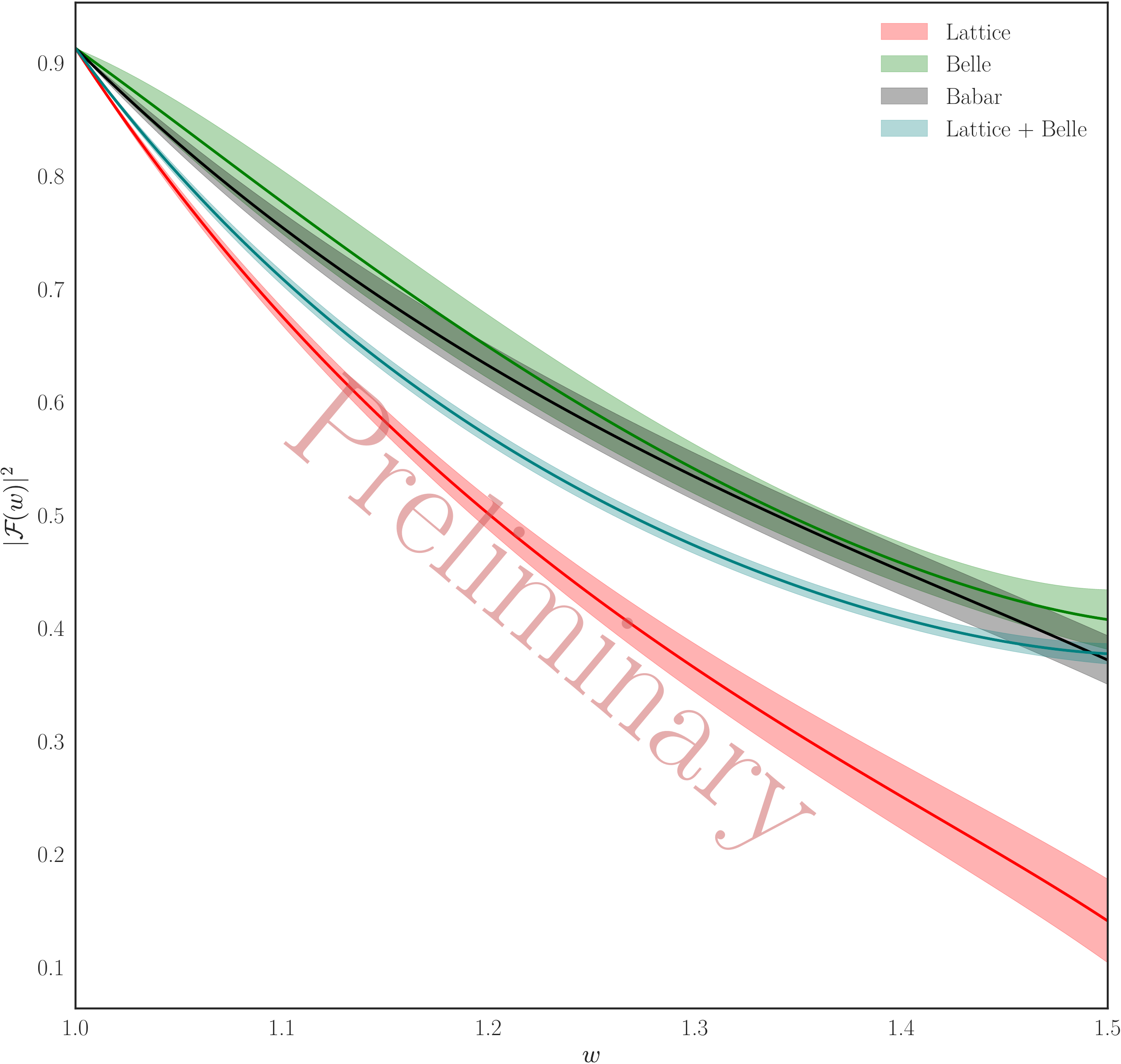}
\caption{Predictions for $|\mathcal{F}|^2$ coming from the lattice, Belle, Babar and the joint fit lattice + Belle (untagged). All the curves are normalized to a common intercept at $w=1$.\label{AdAmp}}
\end{figure}

\begin{figure}[t]
\centering
\includegraphics[width=80mm]{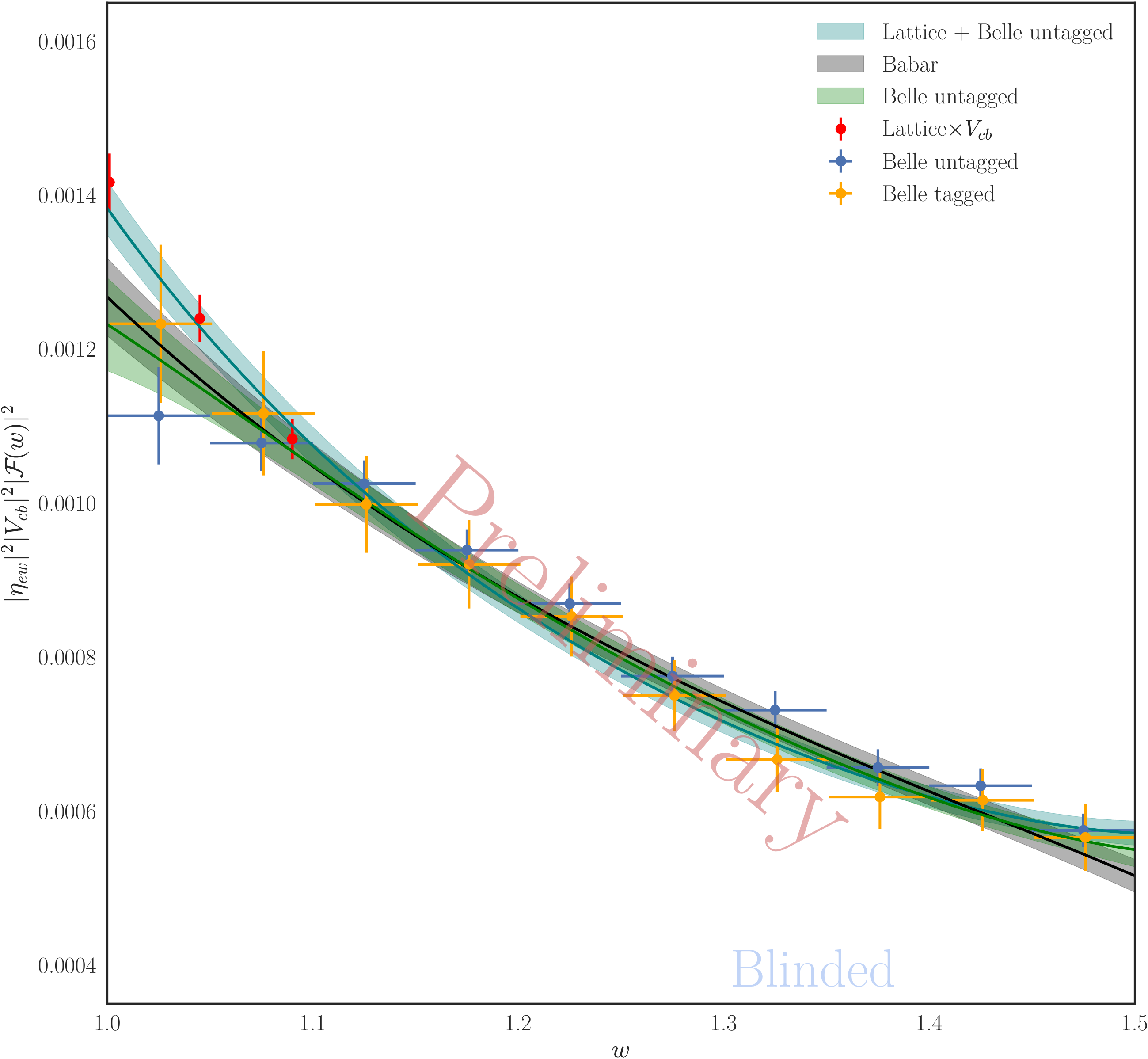}
\caption{Results for $\eta_{\rm{EW}}^2|V_{cb}|^2|\mathcal{F}(w)|^2$ coming from Belle, Babar and the joint fit lattice + Belle (untagged). The $p$~values for the different fits are $p_{\textrm{Belle}}=0.09$
and $p_{\textrm{Joint}}\sim 10^{-6}$. The BaBar collaboration didn't disclose the $p$~value in their article~\cite{Babar}.} \label{fResAll}
\end{figure}

With the lattice data available, one can try a pure lattice prediction of the decay amplitude $|\mathcal{F}(w)|^2$, fitting each form factor to its corresponding $z$~expansion, and compare it with the pure
experimental fits. Experimental data can be fitted only to $\eta_{\rm{EW}}^2|V_{cb}|^2|\mathcal{F}(w)|^2$, but the difference is a proportionality factor, and one can readily compare the shapes of the decay
amplitudes by taking out an overall normalization. This is done in Fig.~\ref{AdAmp}. In Fig.~\ref{fResAll} we show the different predictions for $\eta_{\rm{EW}}^2|V_{cb}|^2|\mathcal{F}(w)|^2$. We include Babar
results, as in \cite{Babar}, our own fit to the Belle untagged dataset and a joint fit using Belle untagged data and the lattice. We also include the data points for both Belle datasets, as well as the lattice
data points multiplied by the $\eta_{\rm{EW}}^2|V_{cb}|^2$ that results from the joint lattice-Belle (untagged) fit. Since the lattice data is blinded, the final value of $\eta_{\rm{EW}}^2|V_{cb}|^2$ coming
from the joint fit is blinded, but other details, such as the slope at low recoil, are not affected and a useful comparison can be done.
The untagged Belle dataset has dramatically reduced the error of the data points. In the low recoil region, we also see a change of curvature that is not observed in the tagged dataset, since the smaller tagged
sample has larger statistical errors. This seems to be at odds with the lattice preliminary prediction, which favors a large slope at small recoil. As a result, although the separate fits have reasonable
$p$~values, the joint fit lattice-Belle (untagged) has a very low $p$~value. In contrast, a joint fit using the Belle tagged dataset and lattice data gives a very similar result but a reasonable $p$~value of
$\gtrsim 0.5$.

One could also use the CLN parametrization for a joint fit with lattice and Belle data. In this case, no matter whether we use the tagged or the untagged dataset, the outcome of the fit is poor. The large
slope at low recoil of the decay amplitude computed with lattice QCD and the shallower behavior at large recoil measured by experiment do not seem to be compatible with the CLN parametrization, which is too
constraining to accommodate the tension in the slope.
\begin{figure}[h]
\centering
\includegraphics[width=80mm]{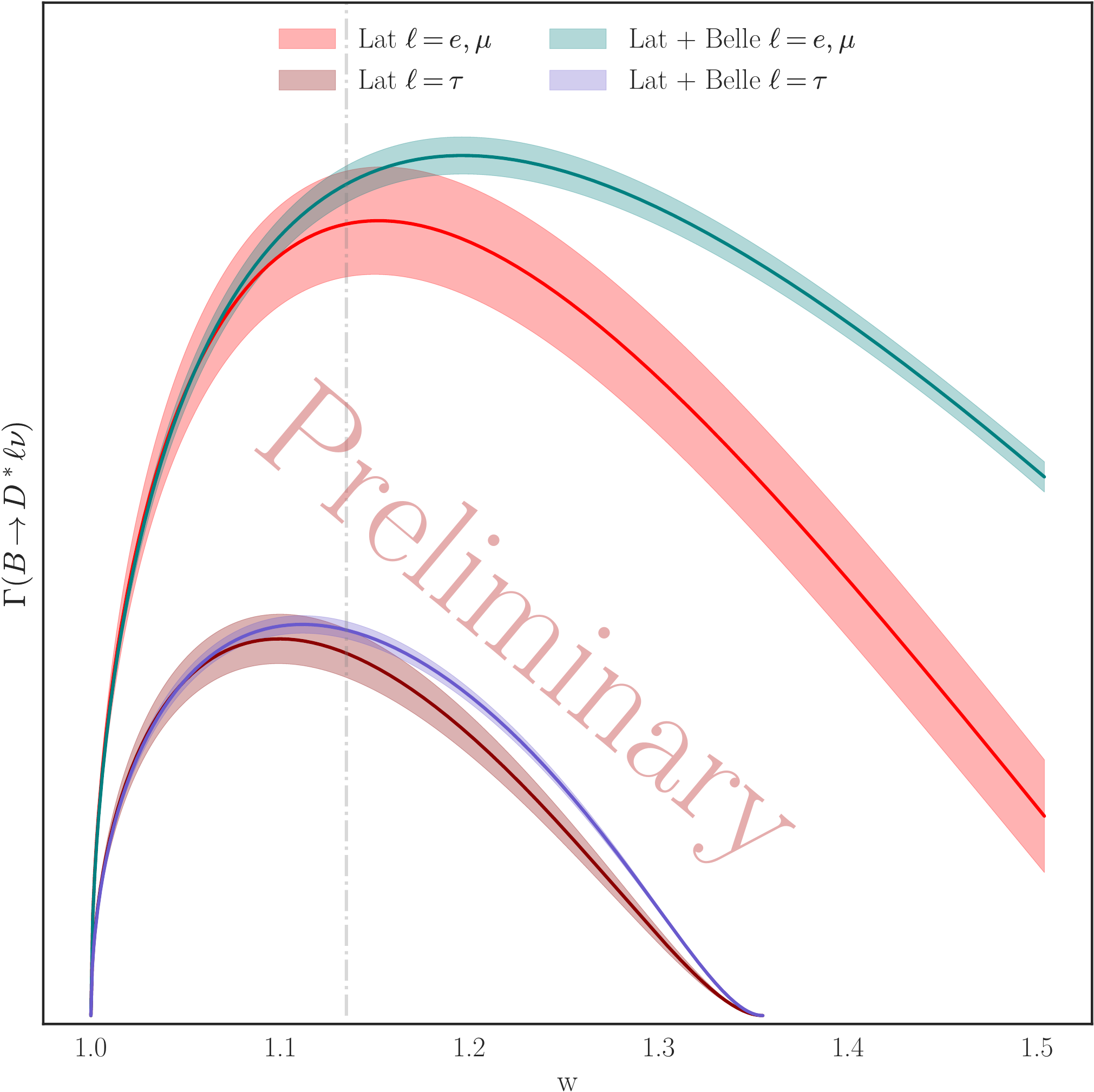}
\caption{Decay rate as a function of $w$ for the lattice QCD calculation (reddish) and for the joint lattice QCD + untagged Belle fit (bluish). The higher set of curves is for massless leptons, whereas the
         lower set of curves is for the $\tau$. The vertical grey line approximately marks the maximum recoil value for which we have lattice data.} \label{dRateInt}
\end{figure}

\begin{figure}[h!]
\centering
\includegraphics[width=80mm]{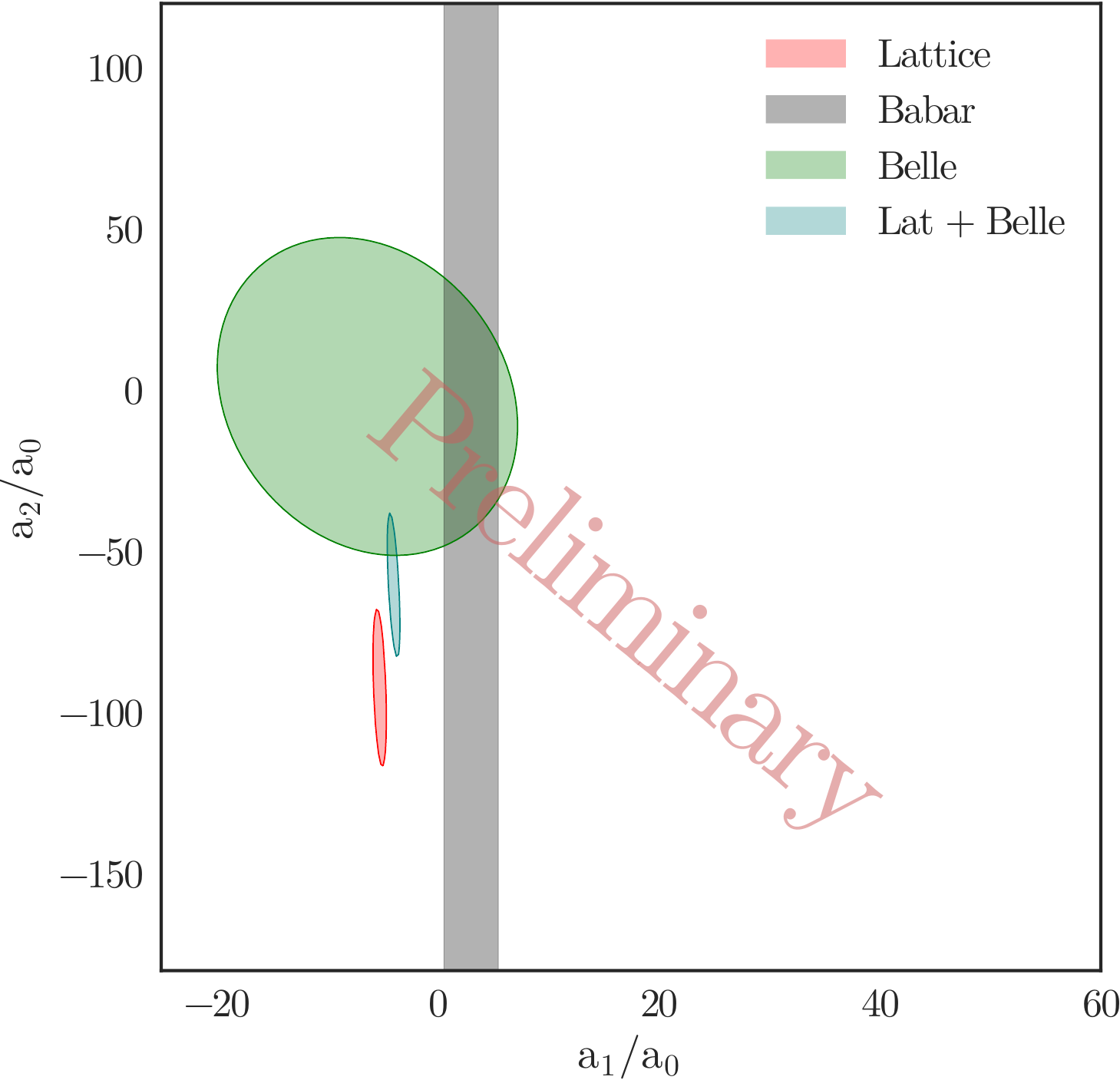}
\caption{One sigma contour plots of the ratio of the coefficient of the $z-$expansion of the $g$ form factor in the BGL parametrization for Belle, Babar, the lattice and the joint fit lattice +
         Belle (untagged). The notation here is $a_j = a_j^g$.} \label{rFits1}
\end{figure}
\begin{figure}[h!]
\centering
\includegraphics[width=80mm]{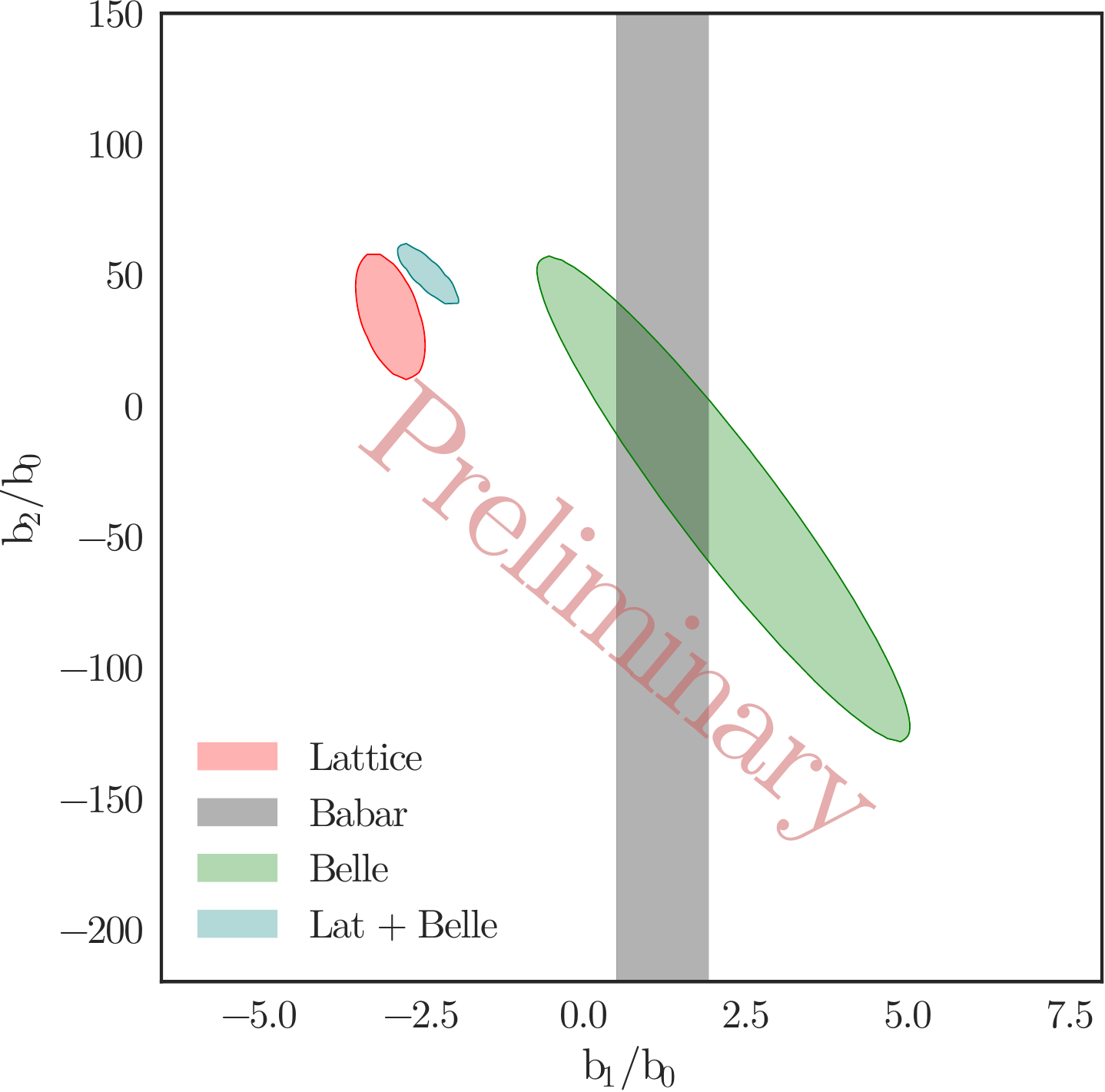}
\caption{One sigma contour plots of the ratio of the coefficient of the $z-$expansion of the $f$ form factor in the BGL parametrization for Belle, Babar, the lattice and the joint fit lattice +
         Belle (untagged). The notation here is $b_j = a_j^f$.} \label{rFits2}
\end{figure}
\begin{figure}[h!]
\centering
\includegraphics[width=80mm]{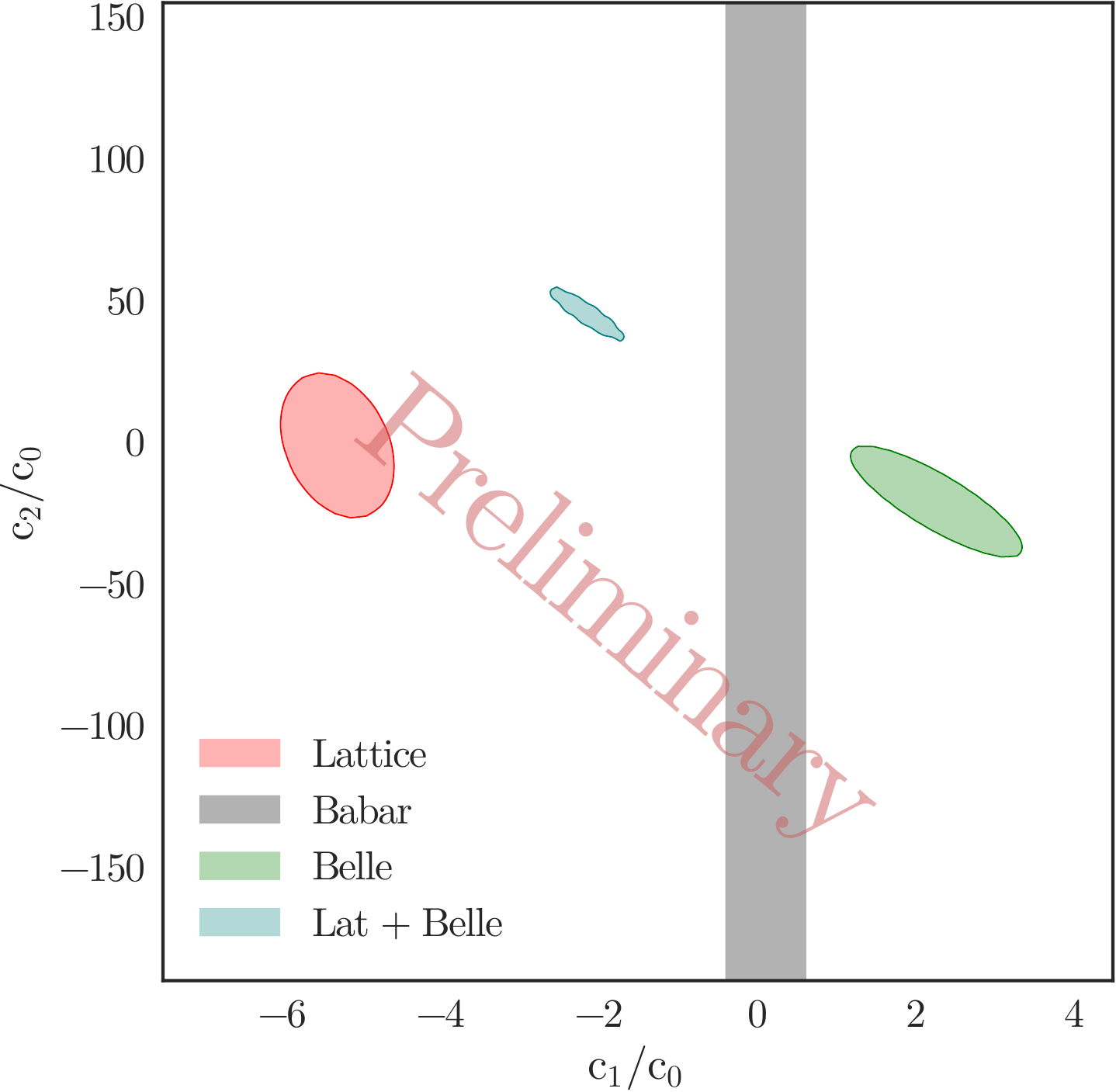}
\caption{One sigma contour plots of the ratio of the coefficient of the $z-$expansion of the $\mathcal{F}_1$ form factor in the BGL parametrization for Belle, Babar, the lattice and the joint fit lattice +
         Belle (untagged). The notation here is $c_j = a_j^{\mathcal{F}_1}$.} \label{rFits3}
\end{figure}
Integrating the different predictions, we can obtain an estimate for $R(D^\ast)$. In Fig.~\ref{dRateInt} we show the decay rate for light leptons and the $\tau$ as predicted from the lattice and from the
joint fit  lattice + Belle. The lattice only data is multiplied by a reasonable value of $\eta_{\rm{EW}}^2|V_{cb}|^2$, so it is properly normalized, although this normalization cancels in the calculation of the
ratio. In the plot we observe a strikingly different behavior in the large recoil region. In that region, the dominant form factor of the BGL parametrization is $\mathcal{F}_1$, the rest are suppressed by
a factor of $q^2$. In Figs.~\ref{rFits1}--\ref{rFits3}~\footnote{Babar uses a different definition of $z$ that slightly modifies the results of the BGL coefficients with respect to the more standard choice of
$z$ as defined in Eq.~(\ref{zParm}). Nonetheless, the differences are quite small and do not affect the general picture exposed here.} we compare the ratios of the $z$~expansion coefficients of the different
fits at one sigma. The fact that we are using ratios allows us to compare the lattice results with experiment, even if the lattice coefficients are blinded. Even though there are some differences in the
coefficients for the $f$ and $g$ BGL form factors, the ratios would agree within a $2\sigma$ ellipsoid. The ratios for $\mathcal{F}_1$, in contrast, show large differences between the pure experimental results
and the results involving lattice data, which explains why the curves in Fig.~\ref{dRateInt} are so different at large recoil. This behavior is currently being studied carefully.

\section{Conclusions}
The flavor-physics community has been facing a long-standing puzzle with the discrepancy between the inclusive and the exclusive determinations of $V_{cb}$. Two years ago, an interesting and elegant proposal
was introduced~\cite{GRINSTEIN2017359, BIGI2017441}, and that seemed to solve the problem. Lattice-QCD calculations were expected to shed light on the issue, but the preliminary lattice results and the latest
experimental fits suggest that the resolution of this puzzle is more complicated than thought.

This situation calls for careful scrutiny of lattice calculations, as well as of the experimental results, in order to extract any conclusions. Hence the situation remains unclear, although with the forthcoming
improvements in both experiment (Belle II) and theory we should have much more information in the coming years to make a better assessment of the puzzle. On the lattice side, very soon we will have results for
two different calculations of the form factors at non-zero recoil. These calculations have very different systematic errors, and an eventual agreement would give reliability to the final result. On the other
hand, the Fermilab/MILC collaboration has a very well established roadmap to improve the precision of its own calculation in three steps. Even if the current status of $V_{cb}$ is a bit uncertain, we have
reason to be optimistic about the future.

\medskip

\begin{acknowledgments}
The author would like to thank Bob Kowaleski and Biplab Dey from the Babar collaboration for useful discussions we held during the FPCP 2019 meeting in Victoria, Canada. Computations for this work were
carried out with resources provided by the USQCD Collaboration, the National Energy Research Scientific Computing Center and the Argonne Leadership Computing Facility, which are funded by the Office of Science
of the U.S. Department of Energy; and with resources provided by the National Institute for Computational Science and the Texas Advanced Computing Center, which are funded through the National Science
Foundation's Teragrid/XSEDE Program. This work was supported in part by the U.S. Department of Energy under grants No. DE-FC02-06ER41446 (C.D.) and No. DE-SC0015655 (A.X.K.), by the U.S. National Science
Foundation under grants PHY10-67881 and PHY14-17805 (J.L.), PHY14-14614 (C.D., A.V.); by the Fermilab Distinguished Scholars program (A.X.K.). Fermilab is operated by Fermi Research Alliance, LLC, under
Contract No. DE-AC02-07CH11359 with the United States Department of Energy, Office of Science, Office of High Energy Physics.
\end{acknowledgments}

\end{document}